\documentclass[aps,prb,twocolumn,showpacs,floatfix,superscriptaddress]{revtex4-1}
\usepackage{graphicx,color}
\usepackage{amsfonts}
\usepackage[figuresright]{rotating}
\usepackage{amssymb}
\usepackage{amsmath}
\usepackage{psfrag}
\usepackage{subfigure}
\usepackage{multirow}
\usepackage{tabularx}
\usepackage{textcomp}
\usepackage{units}
\usepackage{hyperref}
\hypersetup{
 pdfnewwindow=true, colorlinks=true,
 linkcolor=blue, anchorcolor=blue,
 citecolor=blue, filecolor=blue,
 menucolor=blue, urlcolor=blue}

\usepackage{graphicx}
\usepackage{epstopdf}
\usepackage{dcolumn}
\usepackage{bm}
\usepackage{color}%

\begin{document}

\title{Exfoliation energy, quasi-particle bandstructure, and excitonic properties\\of selenium and tellurium atomic chains}
\author{Eesha Andharia}
\email{esandhar@uark.edu}
\affiliation{Department of Physics, University of Arkansas, Fayetteville, AR 72701, USA}
\affiliation{Institute of Nanoscale Science and Engineering, University of Arkansas, Fayetteville, AR 72701, USA}
\author{Thaneshwor P. Kaloni}
\affiliation{Department of Physics, University of Arkansas, Fayetteville, AR 72701, USA}
\affiliation{Institute of Nanoscale Science and Engineering, University of Arkansas, Fayetteville, AR 72701, USA}
\author{Gregory J. Salamo}
\affiliation{Department of Physics, University of Arkansas, Fayetteville, AR 72701, USA}
\affiliation{Institute of Nanoscale Science and Engineering, University of Arkansas, Fayetteville, AR 72701, USA}
\author{Shui-Qing Yu}
\affiliation{Department of Electrical Engineering, University of Arkansas, Fayetteville, AR 72701, USA}
\affiliation{Institute of Nanoscale Science and Engineering, University of Arkansas, Fayetteville, AR 72701, USA}
\author{Hugh O. H. Churchill}
\affiliation{Department of Physics, University of Arkansas, Fayetteville, AR 72701, USA}
\affiliation{Institute of Nanoscale Science and Engineering, University of Arkansas, Fayetteville, AR 72701, USA}
\author{Salvador Barraza-Lopez}
\email{sbarraza@uark.edu}
\affiliation{Department of Physics, University of Arkansas, Fayetteville, AR 72701, USA}
\affiliation{Institute of Nanoscale Science and Engineering, University of Arkansas, Fayetteville, AR 72701, USA}

\begin{abstract}
Effects that are not captured by the generalized-gradient density-functional theory play a prominent effect on the structural binding, and on the electronic and optical properties of reduced-dimensional and weakly-bound materials. Here, we report the exfoliation energy of selenium and tellurium atomic chains with non-empirical van der Waals corrections, and their electronic and optical properties with the GW and Bethe-Salpeter formalisms. The exfoliation energy is found to be within 0.547 to 0.719 eV/u.c. for the selenium atomic chain, and 0.737 to 0.926 eV/u.c. for the tellurium atomic chain (u.c. stands for unit cell), depending on the approximation for the van der Waals interaction and the numerical tool chosen. The GW electronic bandgap turned out to be 5.22--5.47 (4.44--4.59) eV for the Se (Te) atomic chains, with the lowest bound obtained with the Godby-Needs (GB), and the upper bound to the Hybertsen-Louie (HL) plasmon-pole models (PPMs). The binding energy of the ground-state excitonic state ranges between 2.69 to 2.72 eV for selenium chains within the HL and GB PPM, respectively, and turned out to be 2.35 eV for tellurium chains with both approximations. The ground state excitonic wave function is localized within 50 \AA{} along the axis for both types of atomic chains, and its energy lies within the visible spectrum: blue [2.50(GN)--2.78(HL) eV] for selenium, and yellow--green [2.09(GN)--2.28(HL) eV] for tellurium, which could be useful for LED applications in the visible spectrum. 
\end{abstract}
\date{\today}

\maketitle

\section{Introduction}
Several low dimensional semiconductors such as silicon, GaP, and SiGe nanowires,\cite{Algra,Conesa-Boj,Hauge1,RevModPhys.82.427,Zardo,RiccardoNanolett2017} conjugated polymeric chains,\cite{cp1,vdH,Samsonidze} carbon nanotubes,\cite{Ijima} transition metal dichalcogenide nanotubes,\cite{Tenne} and single-walled boron nitride nanotubes,\cite{PhysRevLett.96.126105} exhibit a large excitonic binding energy that originates from the reduced dielectric screening and strong quantum confinement in one dimension.

Selenium and tellurium have historically played a fundamental role in the early validation of current approaches to density-functional theory (DFT):\cite{hohenberg1964inhomogeneous,kohn1965self} from showing that {\em ab initio} local-density total-energy calculations are viable for the study of molecular crystals,\cite{vdb1}
to discussions of 1D materials in (``polymerized'') selenium helical chains,\cite{springborg} to pioneering demonstrations that adding generalized-gradient corrections to the local-density approximation helps in better describing the atomistic structure of bulk Se and Te.\cite{KresseSeTe}  In their bulk and crystalline form, these materials display an atomistic structure of strongly-bonded helical atomic chains with three atoms per unit cell, with a comparatively weaker interaction among neighboring chains\cite{physbook,Olechna,Joannopoulos} that may permit exfoliation down to the single atomic chain limit.

Five computational works have dealt with the structural, electronic, and optical properties of these helical chains: Olechna and Knox provided a tight-binding bandstructure study of selenium chains;\cite{Olechna} Springborg and Jones carried out DFT studies within LDA\cite{ceperley,zunger} in an in-house code;\cite{springborg} there are two publications by Waghmare and coworkers\cite{kahaly2008size,PhysRevB.75.245437} where LDA and PBE\cite{perdew1996generalized} exchange-correlation functionals and norm conserving (NC) pseudopotentials\cite{rappe} as implemented in the {\em ABINIT} code\cite{abinit} were employed; and work by Tuttle, Alhassan and Pantelides\cite{nano7050115} done with the {\em VASP} code,\cite{VASP} where the semi-empirical van der Waals (vdW) correction due to Grimme\cite{grimme} was added to the PBE functional, and PAW pseudopotentials\cite{paw,pawvasp} were employed. The spacing among periodic chains (10 \AA) is disclosed only in two of these works.\cite{PhysRevB.75.245437,nano7050115} There is no systematic study of the mechanical and opto-electronic properties of Se and Te helical atomic chains to date that includes van der Waals corrections at the {\em ab-initio} level (e.g., as implemented in Refs.~\onlinecite{thonhauser2007van,langreth2009density,thonhauser2015spin} and \onlinecite{berland2015van} among others). Calculations of these helical chains that include many-body interactions within the GW\cite{Hedin,hl1,hl2,hybertsen1986electron,GWreview} and Bethe-Salpeter\cite{BS} approaches are scarce too. For example, and although not critical for structural properties, employing a spacing of only 10 \AA\ among periodic images, and only eight bands to express the dielectric screening\cite{nano7050115} may lead to dielectric properties that differ from those of a truly isolated atomic chain, in turn impacting predictions of GW-corrected bandstructures and excitonic properties that prompt a revision of these results. Reassessing the exfoliation energy and the optoelectronic properties of these chains is important now, given that the search for Se- or Te-based monoatomic semiconducting nanowires and/or atomically-thin chains is under experimental investigation.\cite{an2003large,gates2002synthesis,Peide,Peide2,Churchill2017}

Towards this goal, we undertake an analysis of the exfoliation energy with ({\em Quantum ESPRESSO} ({\em QE})\cite{giannozzi2009quantum} using two non-empirical vdW corrections. Afterwards, we provide the reenormalized bandstructures of these chains within the GW approximation, and the excitonic absorption spectra and ground state excitonic wavefunction within the Bethe-Salpeter approach, as implemented in the {\em BerkeleyGW} code.\cite{DESLIPPE20121269}

\section{Computational details}
PBE-DFT\cite{perdew1996generalized} calculations were performed with the {\em QE} computer package. In these runs, a hexagonal unit cell with $a=b\ne c$ was employed for bulk samples, and a tetragonal box with $a=b\ne c_{chain}$ was utilized for the calculations involving chains (a graphical depiction of these chains will be provided later on).

In {\em QE} runs, NC\cite{rappe} and PAW\cite{pawgarrity} pseudopotentials were employed. Additional calculations with non-empirical vdW exchange-correlation functionals were performed as well.\cite{thonhauser2007van,thonhauser2015spin,berland2015van,langreth2009density} In optimizing these structures, a force convergence criteria of $10^{-3}$ eV/\AA\ was utilized. The effect of spin-orbit coupling (SOC) was studied too.

Convergence of the total energy was tested against the number of $k-$points and the energy cutoff. In the case of atomic chains, convergence of the total energy and the amount of electronic charge in the vacuum region {\em versus} the in-plane lattice constant was investigated too. In both codes, convergence is reached with a $12\times 12\times 12$ $k-$point grid ($1\times 1\times 20$ in the case of the chains), and within an energy cutoff of 612 eV (45 Ry). Only in the case of tellurium with non-conserving pseudopotentials, the energy cutoff converged at 1100 eV (85 Ry). The converged $k-$point grids and energy cutoffs employed in our DFT calculations are larger than those employed before.\cite{kahaly2008size,PhysRevB.75.245437} Convergence of the total energy, and the amount of charge against the vacuum among the periodic chains --which is important for a proper description of the dielectric environment-- will be discussed later on.

Quasi-particle energies that include self-energy corrections $\Sigma$ can be obtained following the procedure established by Hedin and Lundqvist:\cite{hl2,GWreview}
\begin{eqnarray}\label{eq:eq1}
(T+V_{ion}+V_{H})\Psi_{n,{\mathbf{k}}}(\mathbf{r})+&\int d^3r'\Sigma(\mathbf{r},\mathbf{r}';E_{n,{\mathbf{k}}}^{QP})\Psi_{{n,{\mathbf{k}}}}(\mathbf{r}') \nonumber{}\\
=&E_{n,{\mathbf{k}}}^{QP}\Psi_{n,{\mathbf{k}}}(\mathbf{r}),
\end{eqnarray}
where $T$ is the kinetic energy operator, $V_{ion}$ is the external potential due to atomic nuclei, $V_H$ is the average Coulomb (Hartree) potential due to electrons, and $\Sigma$ is the non-local electron self-energy.

Excitonic properties are in turn obtained by solving Bethe-Salpeter equation for the electron-hole state:\cite{BS,onida,rohlfing2000electron}
\begin{eqnarray}
(E_{c,\mathbf{k}+\mathbf{q}}^{QP}-E_{v,\mathbf{k}}^{QP})A_{vc,\mathbf{k}}^i+\nonumber\\
\int_{BZ}d^3k'\sum_{v',c'}\langle vc,\mathbf{k}|K^{eh}|v'c',\mathbf{k}'\rangle A_{v'c',\mathbf{k}'}^i = \Omega_iA_{vc,\mathbf{k}}^i,
\end{eqnarray}
where $E_{c,\mathbf{k}+\mathbf{q}}^{QP}$ and $E_{v,\mathbf{k}}^{QP}$ are quasiparticle energies at the conduction and valence bands, respectively, $K^{eh}$ is the electron-hole interaction kernel, $\Omega_i$ are the exciton eigenvalues, and $A^i_{vck}$ is the amplitude of the exciton wave function $|i\rangle$:
\begin{equation}\label{eq:3}
|i\rangle = \sum_v\sum_c\sum_{\mathbf{k}}A^i_{vc,\mathbf{k}}|vc,\mathbf{k}\rangle,
\end{equation}
where $v$ labels the quasi-hole and $c$ the quasi-electron states, respectively. The electron-hole interaction kernel $K^{eh}$ consists of a direct term $K^d$, and an indirect repulsive term $K^x$.

\begin{table}[tb]
\caption{Lattice parameters $c$ and $a$, $c/a$ ratio for bulk Se and Te, length of the chain's unit cell $c_{chain}$, DFT bandgaps for these chains $E_{g,chain}$,  and exfoliation energy $E_{exf}$.}\label{ta:ta1}
  \begin{tabular}{|c||c|c||c|c||c|}
    \hline
    &  \multicolumn{2}{c||}{PBE} & \multicolumn{2}{c||}{vdW} & \multicolumn{1}{c|}{}\\
       & (NC) & (PAW) & (NC)& (PAW)  &   Expt.\cite{PhysRevB.16.4404}   \\
    \hline
    \hline
    $a$ (\AA)  &4.516& 4.510              & 4.279& 4.531& 4.368\\
    \hline
    $c$ (\AA)  &5.088& 5.057              & 5.103& 5.229& 4.958\\
    \hline
    $c/a$      &1.127& 1.121              & 1.192& 1.154& 1.135\\
    \hline
    \hline
    $c_{chain}$ (\AA)  &4.970 & 4.949     & 4.957 & 4.952 & -- \\
    \hline
    $E_{g,chain}$ (eV) &2.073 & 2.043     & 2.070 & 1.922 & -- \\
    \hline
    $E_{exf}$ (eV)    &0.152& 0.152       & 0.547 & 0.719 & -- \\
    \hline
    \hline
    \hline
    &  \multicolumn{2}{c||}{PBE} & \multicolumn{2}{c||}{vdW} & \multicolumn{1}{c|}{}\\
      & (NC) & (PAW)  & (NC)& (PAW) &     Expt.\cite{PhysRevB.16.4404}\\
    \hline
    \hline
    $a$ (\AA)&4.454&4.498&              4.351&4.446&  4.451\\
    \hline
    $c$ (\AA)&5.947&5.961&              5.969&6.112&  5.926 \\
    \hline
    $c/a$  &1.335&1.325&                1.372&1.374&  1.331 \\
    \hline
    \hline
    $c_{chain}$ (\AA) &5.663&5.651&     5.599&5.654 &--  \\
    \hline
    $E_{g,chain}$ (eV)&1.720  &1.750&   1.760& 1.551 & --   \\
    \hline
    $E_{exf}$ (eV)   &0.544&0.581&      0.926&0.737 &-- \\
    \hline
\end{tabular}
\end{table}

Using the Kohn-Sham energy eigenvalues obtained form {\em QE} calculations as input, electron-correlation effects were evaluated with the {\em BerkeleyGW} code,\cite{DESLIPPE20121269} employing a $1\times 1 \times 30$ $k-$point mesh. Due to compatibility issues, NC pseudopotentials\cite{rappe}  were employed (which preclude calculation of spin-orbit coupling in the version of {\em BerkeleyGW} we used), and vdW interactions were taken into account via the vdW-DF2 functional.\cite{thonhauser2007van,thonhauser2015spin,berland2015van,langreth2009density} Given that the length of these chain's unit cells is $\sim 5$ \AA, roughly twice the length of an $(n,n)$ carbon nanotube, our choice of $k-$points is equivalent to the $1\times 1 \times 64$ $k-$point mesh employed to study $(n,n)$ carbon nanotubes in Reference~\onlinecite{spartaru}. We verified that a cut-off of 272 eV (20 Ry) was sufficient to reach a converged dielectric matrix, by looking at the convergence of the GW bandgap versus this parameter (more on this later).

A cell wire truncation for the Coulomb interaction among periodic images was employed as well.\cite{beigiprb} We find that 99.996\% or more of the electronic charge was confined to within a radius of $a/4$ from the Se and Te wires' center of mass when $a\ge 20$ \AA. Convergence of the dielectric function, of the GW bandgap, and of the excitonic wavefunctions was ensured by employing 45 unoccupied bands, which is an order of magnitude larger than the four unoccupied bands used in Ref.~\onlinecite{nano7050115}. The absorption spectrum with and without electron-hole interactions was also calculated to obtain the ground-state exciton binding energy.

\section{Results}
\begin{figure}[tb]
\includegraphics[width=0.48\textwidth]{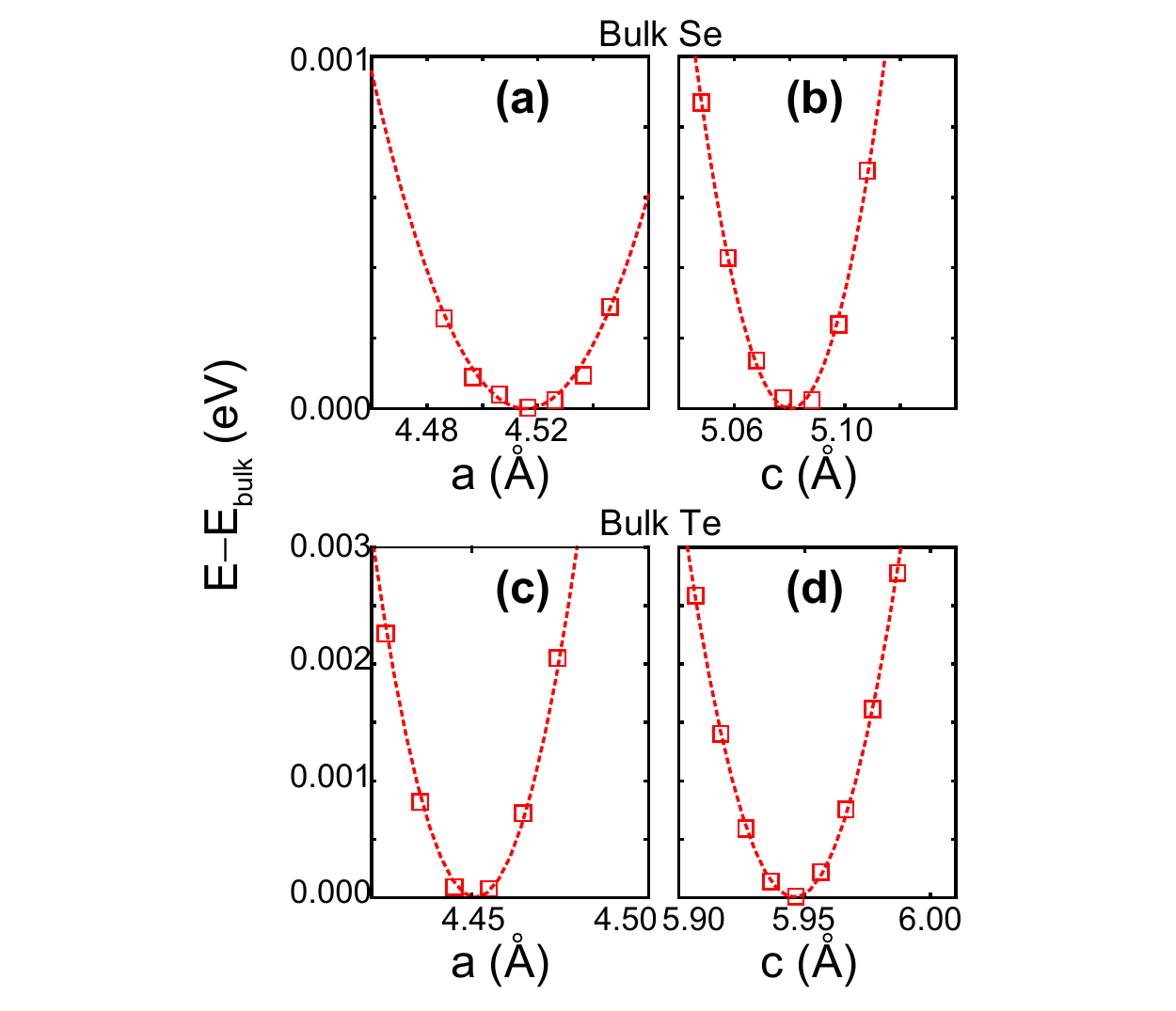}
\caption{Energy $E$ {\em versus} in-plane and out-of-plane lattice constants $a$ and $c$ for (a,b) bulk selenium and (d,e) bulk tellurium, as obtained from the {\em QE} code. The energy is reported with respect to the minimum for each calculation, labeled $E_{bulk}$.}
\label{fig:fig_0}
\end{figure}

The experimental lattice parameters of bulk selenium in a hexagonal unit cell under standard temperature and pressure conditions are \textit{a = b =} 4.366--4.368 \AA, and \textit{c =} 4.955--4.958 \AA.\cite{seexpt1,Ren,PhysRevB.16.4404} The lattice parameters for bulk tellurium are \textit{a = b =} 4.451 \AA, and \textit{c =} 5.926 \AA.\cite{PhysRevB.16.4404} In Fig.~\ref{fig:fig_0}, we display the total energy {\em versus} the in-plane ($a$) and out-of-plane ($c$) lattice parameters as obtained with the {\em QE} code. In these plots, the energy is referred with respect to its minimum magnitude $E_{bulk}$ for a given pseudopotential (NC or PAW) employed.

Structural parameters for bulk samples, the length of the chain on the unit cell $c_{chain}$ for the helical atomic chains, their DFT band gaps $E_{g,chain}$, as well as their exfoliation energy $E_{exf}$, are listed in Table \ref{ta:ta1}.

\begin{figure*}[tb]
\includegraphics[width=0.96\textwidth]{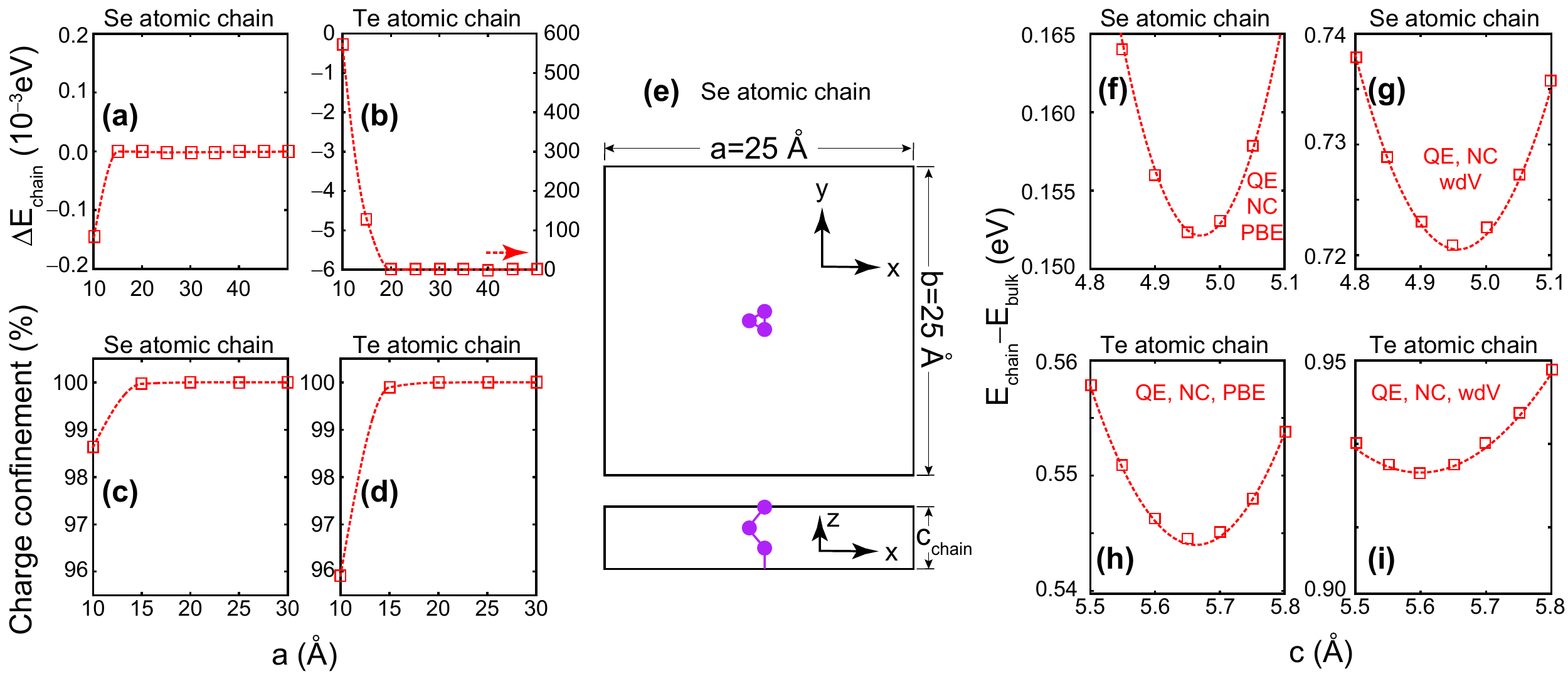}
\caption{(a,b) Convergence of the total energy {\em versus} $a$ on the tetragonal unit cell. (c,d) Total amount of charge $q$ within a cylinder centered about the chain axis with radius $a/4$ {\em versus} $a$ on the tetragonal unit cell. The vertical axes, named charge confinement, display $\frac{q}{18 \text{ electrons}}\times 100$. (e) Atomistic structure of the selenium atomic chain (the structure of the tellurium chain is similar, and not shown for that reason). (f,g) Total energy for the selenium atomic chain without and with vdW corrections. (h,i) Total energy for the tellurium atomic chain without and with vdW corrections, respectively. The minima in subplots (f-i) are at coordinates ($c_{chain},E_{exf}$) listed in Table \ref{ta:ta1}.}
\label{fig:fig_1}
\end{figure*}

As seen in Figs.~\ref{fig:fig_1}(a) and \ref{fig:fig_1}(b), the structural energy of the atomic chain, $E_{chain}$, is slightly sensitive on the magnitude of $a$; this is even more so for the tellurium wire. There, $\Delta E_{chain}=E_{chain}(a)-E_{chain}(50$ \AA{}$)$. The structural energy is still not converged at $a=10$ \AA.\cite{PhysRevB.75.245437,nano7050115} In Figures \ref{fig:fig_1}(c) and \ref{fig:fig_1}(d), the electronic charge is integrated over an $a/4$ radius, measured with respect to the chain's axis of symmetry, and displayed as a percent of the total charge. This helps in understanding the degree of localization of the electronic charge within these chains. Upon comparison, the tellurium chain spills its electronic charge over a larger radius. Figures \ref{fig:fig_1}(c) and \ref{fig:fig_1}(d) indicate that the charge is 99.996 \% contained within a cylinder of $a/4$ radius only when $a = 20$ \AA. As illustrated in Fig.~\ref{fig:fig_1}(e), we actually employed $a=25$ \AA{} for all calculations (DFT, GW, and Bethe-Salpeter) involving chains.

The lattice parameter $c_{chain}$ of the tetragonal cell was subsequently optimized while keeping $a=25$ \AA.  The values of $c_{chain}$ agree within the second decimal digit in all entries in Table \ref{ta:ta1}. As seen in that Table, $c_{chain}$ decreases by about $2.9\pm 1.1$ percent with respect to its value $c$ in bulk selenium, and $5.4\pm 1.1$ percent when compared to $c$ in bulk tellurium, so that these hexagonal chains compress under isolation.

The atomistic structure of these chains is not significantly altered upon inclusion of vdW corrections. Nevertheless, these corrections naturally alter the estimate of binding energy; {\em i.e.}, the ease to exfoliate a single atomic chain. The continuous trends in Figs.~\ref{fig:fig_1}(f-i) are quadratic fits to the data shown in squares. The coordinates at the minimum points, i.e., ($c_{chain},E_{exf}$), are listed in Table \ref{ta:ta1} explicitly.

Without vdW corrections, $E_{exf}=0.152\pm 0.001$ eV/u.c. for the selenium chain, and $0.537 \pm 0.047$ eV/u.c. for the tellurium chain. The vdW functional corrects the underestimation in binding of the PBE functional, and the exfoliation energy obtained with non-empirical vdW corrections turns out to be $0.678\pm 0.116$ eV/u.c. for the selenium chain, and $0.826\pm 0.095$ eV/u.c. for the tellurium one for a 22\% increase. Such energetics imply that it may be similarly likely to exfoliate chains of either element from defect-free, crystalline samples.

Recalling that the chain's unit cell contains three atoms (Fig.~\ref{fig:fig_1}(e)), the exfoliation energies reported in Ref.~\onlinecite{nano7050115} (0.600 eV/u.c, and 0.810 eV/u.c. for the selenium and tellurium atomic chain, respectively) agree with the values presented here. The reader must note, nevertheless, a procedural difference in these estimates. Here, we use the vdW functional for both bulk and chain, while previous work\cite{nano7050115} used an empirical vdW correction\cite{grimme} and the exfoliation energy was computed as the difference between an energy obtained with the PBE functional (chain), and an energy with empirical vdW corrections (bulk).

\begin{figure}[tb]
\includegraphics[width=0.48\textwidth]{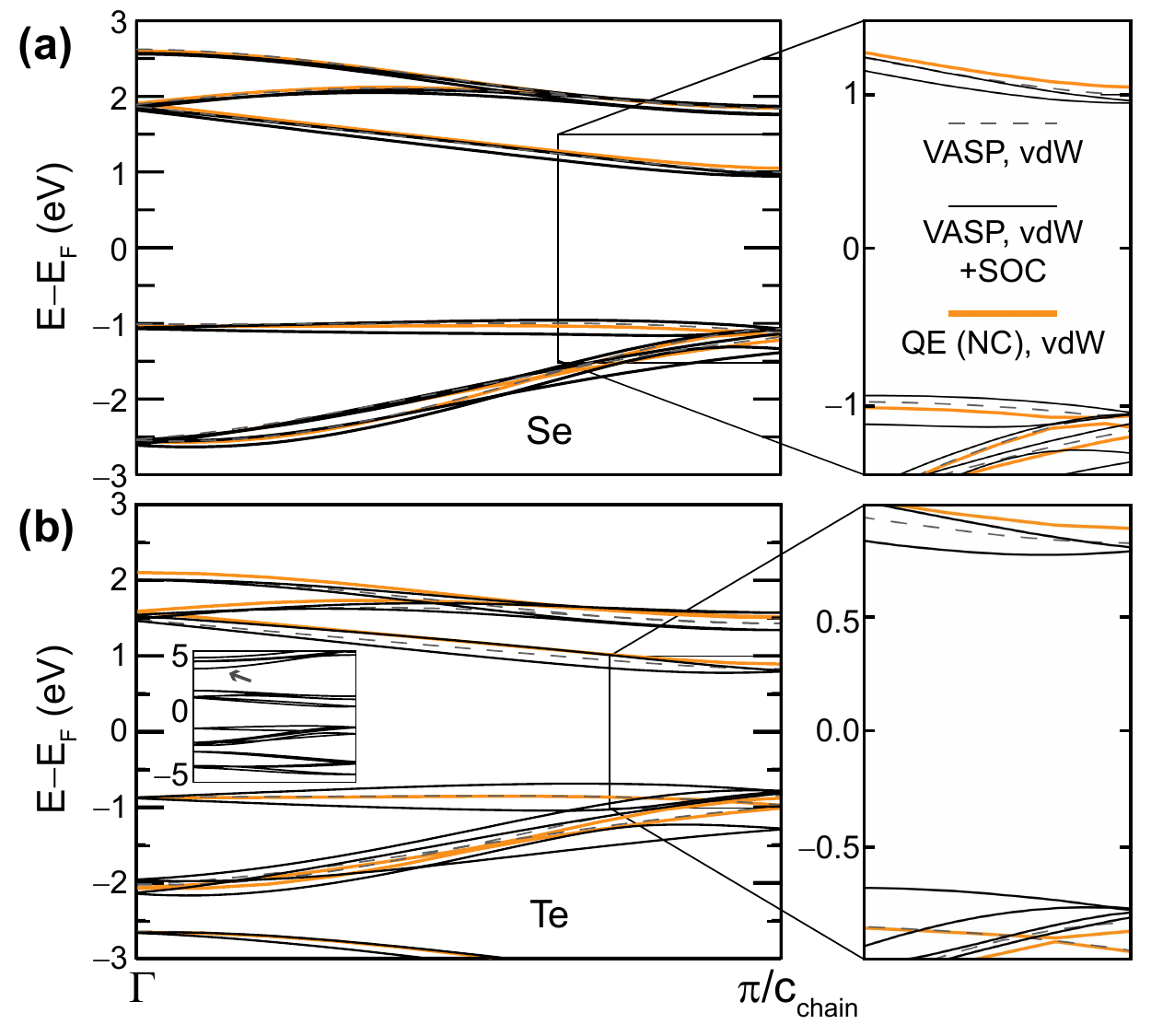}
\caption{Bandstructures for (a) selenium and (b) tellurium atomic chains with vdW exchange-correlation functionals. The golden lines correspond to a calculation with the {\em QE} code with NC pseudopotentials; the dashed line to a calculation with the {\em VASP} code without spin-orbit coupling; and the solid black lines to a {\em VASP} calculation with spin-orbit coupling. Inset in (b) shows bands above 3 eV.}
\label{fig:fig_2}
\end{figure}

Bandstructures computed along the $\Gamma-\pi/c_{chain}$ line obtained with the {\em QE} (NC) and {\em VASP} codes that include vdW corrections are plotted in Fig.~\ref{fig:fig_2}. The golden lines were obtained with {\em QE} (NC); the dashed line with {\em VASP}; and the solid lines correspond to a {\em VASP} calculation with the SOC turned on. Zoom-ins highlight the bandgap near the zone edge around the $\pi/c_{chain}$ point. In addition, the inset in Fig.~\ref{fig:fig_2}(b) showcases bands over a larger energy range.

  In Figure \ref{fig:fig_2}, the selenium atomic chain displays a bandgap of 2.070 (2.076) eV according to our well-converged {\em QE} ({\em VASP}) calculations.
As seen in Table \ref{ta:ta1}, the bandgap is only marginally modified when PBE calculations --without vdW corrections but still on fully relaxed structures-- are performed. This bandgap is 14\% larger than the one obtained with the {\em ABINIT} code (1.82 eV), likely due to the potentially different optimization of the NC pseudopotentials employed previously,\cite{kahaly2008size} that also leads to a $\sim 2$\% more elongated chain ($c_{chain}=5.079$ \AA) in comparison to the six smaller and nearly identical values of $c_{chain}$ (4.949--4.970) we find in our calculations and those in Ref.~\onlinecite{nano7050115}.

\begin{figure}[tb]
\includegraphics[width=0.48\textwidth]{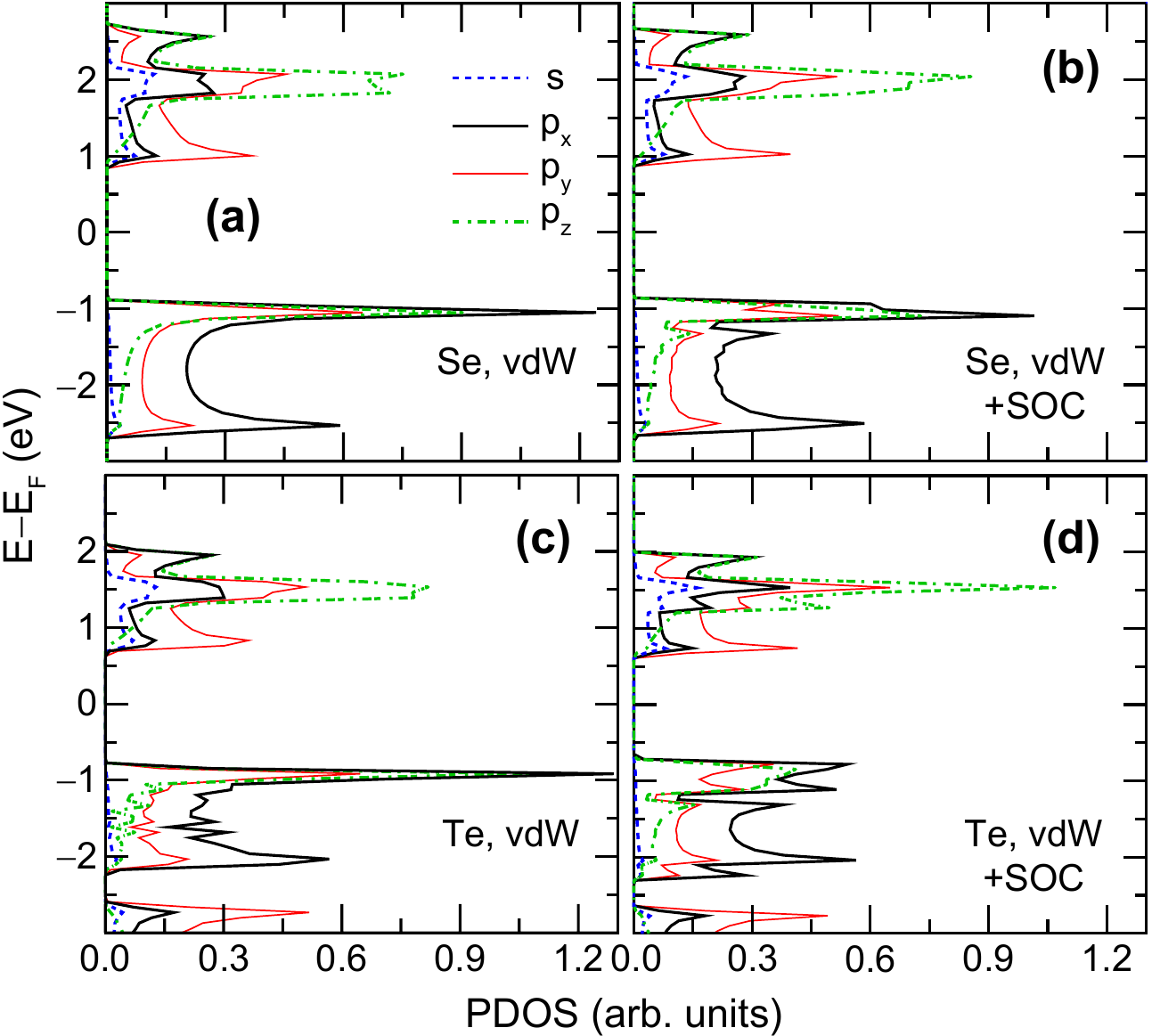}
\caption{Orbital-resolved density of states (DOS) for (a-b) selenium and (c-d) tellurium atomic chains with and without SOC ({\em VASP} code employed). Overall, the height of the van Hove singularities does not change drastically upon inclusion of SOC corrections, but SOC does produce a considerable split near the valence band, especially for the tellurium chain, and an increase on DOS at about 2.0 (1.5) eV for the Se (Te) chain.}
\label{fig:fig_3}
\end{figure}

The bandgap reported before (1.64 eV) for the tellurium wire, obtained with PBE-DFT and without SOC,\cite{PhysRevB.75.245437} is also (9\%) smaller than the ones we found with PBE without the SOC turned on (1.720--1.787 eV). When vdW corrections are turned on, and still without SOC, a {\em QE} calculation with PAW pseudopotentials gives a gap of 1.551 eV, while its value is 1.675 eV with {\em VASP}, and 1.760 with {\em QE} NC.

As the SOC is turned on (solid black lines in Fig.~\ref{fig:fig_2}), the spin-degenerate bands become split. The splitting becomes stronger for the valence bands for both chains, as evidenced by the DOS shown in Fig.~\ref{fig:fig_3}, which was obtained on a $1\times 1\times 80$ $k-$point grid, with a gaussian smearing of 0.050 eV, and an energy resolution of 0.075 eV.
The valence band has a predominant $p_x$, $p_z$ character, while the conduction band has a larger contribution from the $p_y$ orbital. In Fig.~\ref{fig:fig_3}, the height of the van Hove peak near the valence band edge gets reduced as SOC is turned on; compare Fig.~\ref{fig:fig_3}(a) {\em versus} \ref{fig:fig_3}(b); and Fig.~\ref{fig:fig_3}(d) {\em versus} \ref{fig:fig_3}(d). The smaller height of the van Hove singularity at the top of the valence band is more so for the case of tellurium because it has a stronger atomic mass and therefore a stronger SOC. Two additional features arising from SOC are a split DOS near $-2$ eV in Fig.~\ref{fig:fig_3}(d), and the higher peaks at about 2.0 (1.5) eV for the selenium (tellurium) chain, which results from four bands meeting at the $\Gamma-$point (due to the initial spin degeneracy) that accidentally overlap over a larger segment of $k-space$ once split due to SOC and thus generate a slightly higher DOS feature at these energies.

Upon inclusion of SOC, the bandgap reduces by 0.058 eV for the selenium chain, and 0.095 eV for the tellurium chain. Spin-orbit coupling is a relativistic property that can be further tuned by crystal symmetry. For any given band, its intrinsic band splitting due to SOC is not expected to be greatly altered upon inclusion of GW corrections (we do not know of any work where the opposite has been claimed).

\begin{figure}[tb]
\includegraphics[width=0.48\textwidth]{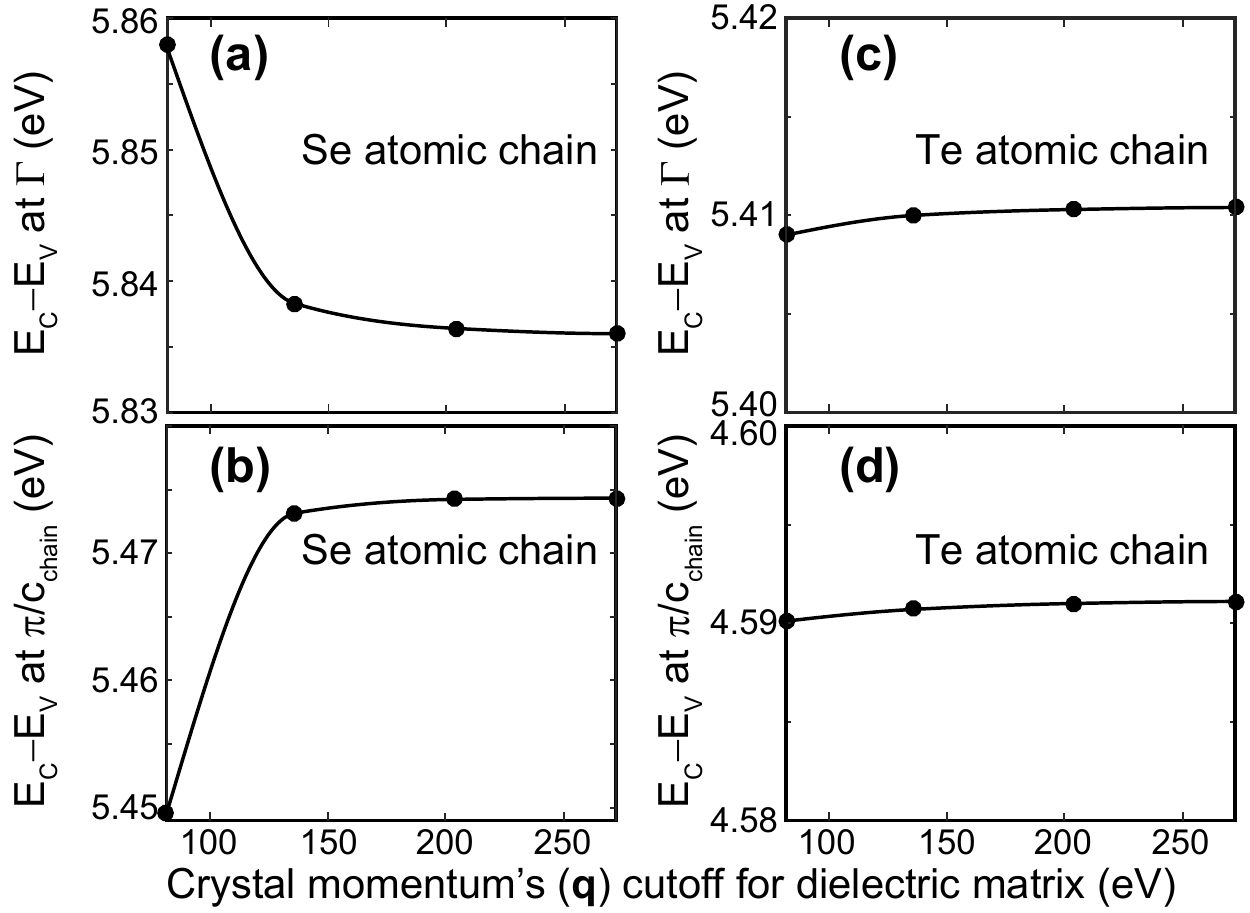}
\caption{Convergence of the electronic bandgap (subplots (a) and (c)) and of the energy difference at the end of the first Brillouin zone, point $\pi/c_{chain}$ (subplots (d) and (e)), {\em versus} the energy cutoff for the $k-$points within the {\em BerkeleyGW} code, for the selenium and tellurium atomic chains, with the HL approximation to the plasmon-pole model (PPM).}
\label{fig:fig_4}
\end{figure}

Quasiparticle energies $E_{n,\mathbf{k}}^{QP}$ were next obtained within the $G_0W_0$ approximation on the simulation box having $a=25$ \AA, using 30 $k-$ points to produce mean-field inputs, including all $18/2=9$ occupied bands (our pseudopotentials include two $s-$ electrons, and four $p-$electrons in the valence), and 45 unoccupied bands for convergence. With this choice of unoccupied bands, we further show in Fig.~\ref{fig:fig_4} the convergence of the $\mathbf{q}-$grid employed to construct the dielectric matrix within the HL approach, by means of the evolution of the bandgap at the $\Gamma$ and the $\pi/c_{chain}$ $k-$points. Although convergence was reached at 200 eV, we settled for a 270 eV cutoff. We also present results within the Godby-Needs\cite{Godby,Godby2} approach to the plasmon-pole model with a $q-$point grid cutoff of 200 eV (according to Reference \onlinecite{WuGoldby}, convergence is reached on a smaller grid within the GN approach, hence our choice). In the GN approach, the frequency integration in the expression for the self-energy is carried out along the
imaginary axis, which is advantageous since the pole structure along the real axis is avoided.\cite{GWreview}

\begin{figure}[tb]
\includegraphics[width=0.48\textwidth]{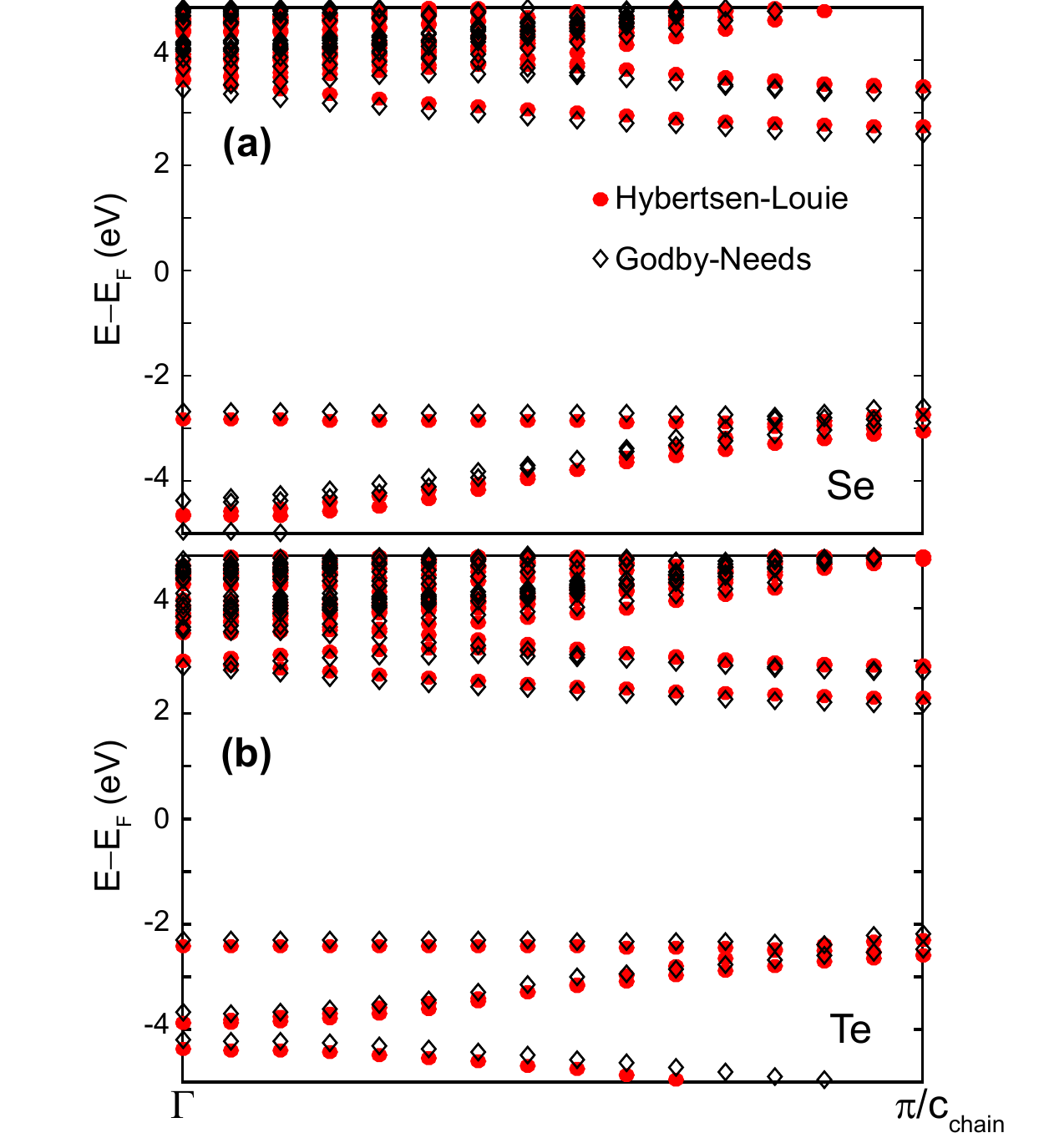}
\caption{$G_0W_0$-corrected bandstructure of (a) selenium and (b) tellurium atomic chains with the HL (full dots) and GN (open diamonds) approximations. The bandgap is direct.}
\label{fig:fig_5}
\end{figure}

As seen in Fig.~\ref{fig:fig_5}, the $G_0W_0$ electronic bandgap turned out to be 5.22--5.47 (4.44--4.59) eV for the Se (Te) atomic chains, with the lowest bound obtained with the GB, and the upper bound to the HL plasmon-pole models. Convergence of all relevant input parameters implies that the small differences observed are due to procedural differences in these two methods. Being a relativistic and crystal-symmetry related effect, in principle barely modified by the dielectric environment, inclusion of SOC (not available in the version of {\em BerkeleyGW} we employed while also including vdW corrections), would probably split the bands to an extent not too dissimilar from that seen in Fig.~\ref{fig:fig_2}.

\begin{figure}[tb]
\includegraphics[width=0.48\textwidth]{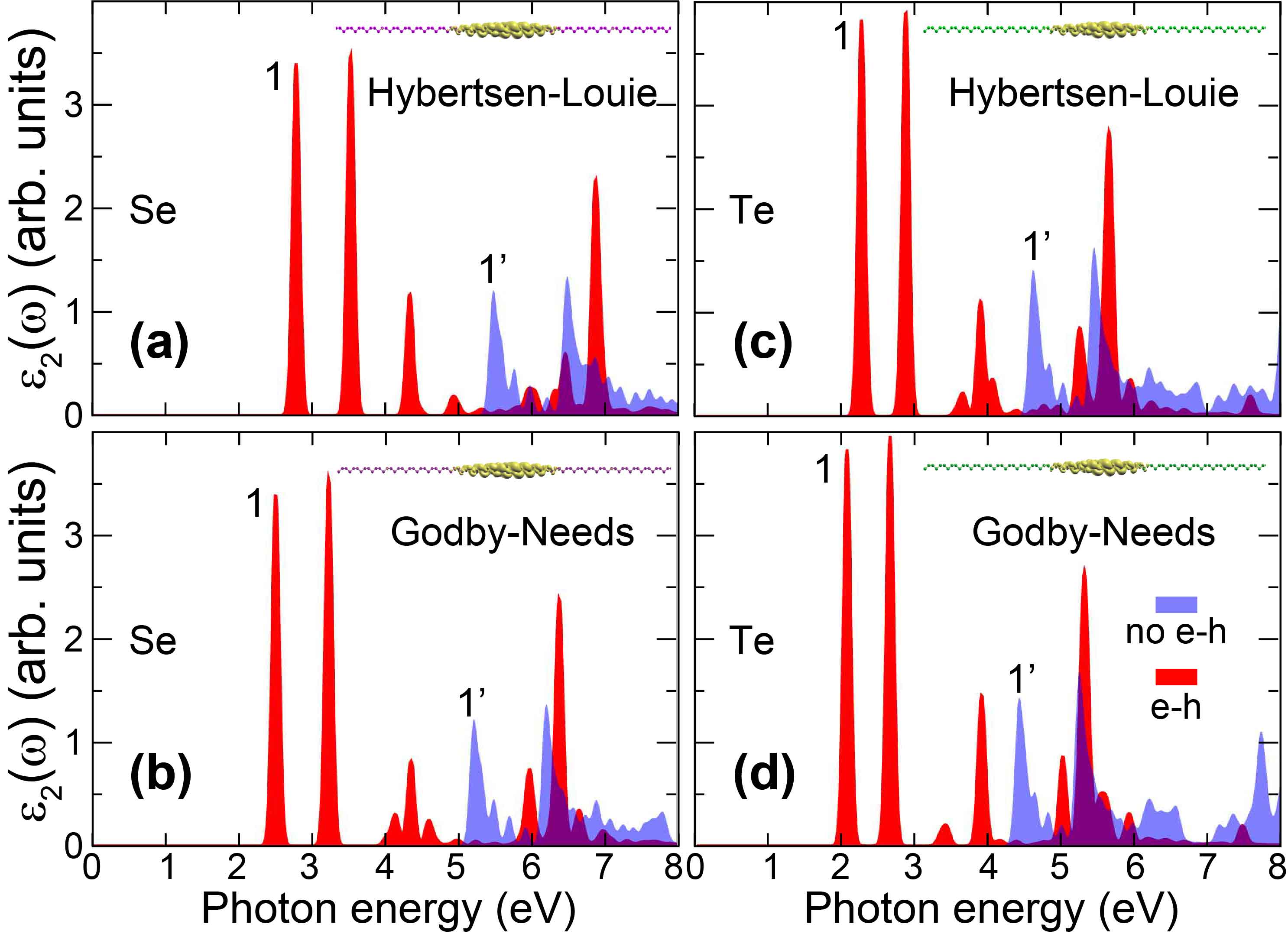}
\caption{Imaginary part of dielectric function (absorption spectrum) with and without electron-hole interactions. The position of the first peaks is highlighted. The blue curve was drawn semi-transparent, which leads to a gray color at energies where the two curves overlap. Insets show the extent ($\sim 50$ \AA) of the excitonic ground state wavefunction.}
\label{fig:fig_6}
\end{figure}

The absorption spectrum (the imaginary part of the dielectric function) is shown in Fig.~\ref{fig:fig_6} for both HL and GN PPMs for selenium and tellurium atomic chains with and without electron-hole interactions. There, the difference in the energy of the peaks position with and without electron-hole interactions provides the excitonic binding energy $\Omega_i$. Peaks listed by 1 and 1' correspond to the ground-state eigenvalues with and without electron-hole interaction. These excitonic energies lie within the visible spectrum: blue (2.50--2.78 eV) for the selenium chain and yellow--green (2.09--2.28 eV) for the tellurium chain, with lower bounds obtained in GN-PPM runs and upper ones obtained with HL-PPM, and are not expected to change drastically upon inclusion of SOC, besides additional splittings of the observed energy features, following arguments provided in previous lines. Nevertheless, spin-selection rules on the SOC-split bands will provide a handle for further experimental control of these excitations.

The binding energy of the ground-state excitonic state $\Omega_0$ ranges between 2.69 to 2.72 eV for selenium chains within the HL and GB PPM, respectively, and turned out to be 2.35 eV for tellurium chains with both approximations. Insets in Fig.~\ref{fig:fig_6} display the degree of localization of the ground state excitonic wave function along the wire, which occurs within 50 \AA{} along the axis for both types of atomic chains.

The structural properties and exfoliation energy we found agrees with those listed in Ref.~\onlinecite{nano7050115}. Nevertheless, the $G_0W_0$ bandgaps listed there (3.0 and 2.4 eV) are 50\% smaller than the ones we found. We also note that the $G_0W_0$-corrected bandstructure in that work misses all states around 4 eV seen in Fig.~\ref{fig:fig_5}, in which 45 unoccupied bands were employed. The ratio of these bandgaps in that paper (2.4/3.0=0.8) happens to be nearly equal to the ratio we found ($4.4/5.2=0.8$), which appears to imply that the small number of unoccupied bands in that work leads to a similar relative underestimation of the dielectric properties for both wires, such that the bandgaps are underestimated by the same percentage with respect to our converged results. (This makes sense, because the limiting case in which the number of empty bands is zero would correspond to no correction of the dielectric environment and would lead to a 2.1 (1.7) eV bandgap. A similar argument of almost zero correction holds when the number of occupied bands is close to zero.

From then on, exciton eigenvalues and exciton binding energies we find are larger than those listed in Ref.~\onlinecite{nano7050115}, in which scarce additional procedural details were listed. Lacking more information, the truncation of sums over only four empty and four occupied bands in that work appear to lead to an underestimated modification of the dielectric environment within GW and Bethe-Salpeter equations, which propagate and explain their  underestimation of their accompanying many-body corrections. No single convergence study was provided in that work.

\section{Conclusions}
We have reported the exfoliation energy of selenium and tellurium atomic chains with non-empirical van der Waals corrections, and their electronic and optical properties with the GW and Bethe-Salpeter formalisms. The exfoliation energy is found to be within 0.547 to 0.719 eV/u.c. for the selenium atomic chain, and 0.737 to 0.926 eV/u.c. for the tellurium atomic chain.

The $G_0W_0$ electronic bandgap turned out to be 5.22--5.47 (4.44--4.59) eV for the Se (Te) atomic chains, with the lowest bound obtained with the Godby-Needs (GB), and the upper bound to the Hybertsen-Louie (HL) plasmon-pole models (PPMs). The binding energy of the ground-state excitonic state ranges between 2.69 to 2.72 eV for selenium chains within the HL and GB PPM, respectively, and turned out to be 2.35 eV for tellurium chains with both approximations.

The ground state excitonic wave function is localized within 50 \AA{} along the axis for both types of atomic chains, and its energy lies within the visible spectrum: blue [2.50(GN)--2.78(HL) eV] for selenium, and yellow--green [2.09(GN)--2.28(HL) eV] for tellurium. The location of these exciton energies within the visible spectrum could be relevant for photovoltaic applications.

\section*{Acknowledgments}
E.A, H.O.H.C and S.B.-L. were funded by the NSF (Grant No. DMR-1610126). Calculations were performed at NERSC ({\em Cori} and {\em Edison}).

%

\end{document}